# Variant N = 1 Supersymmetric Non-Abelian Proca-Stueckelberg Formalism in Four Dimensions

Hitoshi Nishino[1)] and Subhash Rajpoot[2)]

*Department of Physics & Astronomy*
*California State University*
*1250 Bellflower Boulevard*
*Long Beach, CA 90840*

## Abstract

We present a new (variant) formulation of $N=1$ supersymmetric compensator mechanism for an arbitrary non-Abelian group in four dimensions. We call this 'variant supersymmetric non-Abelian Proca-Stueckelberg formalism'. Our field content is economical, consisting only of the two multiplets: (i) A Non-Abelian vector multiplet $(A_\mu{}^I, \lambda^I, C_{\mu\nu\rho}{}^I)$ and (ii) A compensator tensor multiplet $(B_{\mu\nu}{}^I, \chi^I, \varphi^I)$. The index $I$ is for the adjoint representation of a non-Abelian gauge group. The $C_{\mu\nu\rho}{}^I$ is originally an auxiliary field Hodge-dual to the conventional auxiliary field $D^I$. The $\varphi^I$ and $B_{\mu\nu}{}^I$ are compensator fields absorbed respectively into the longitudinal components of $A_\mu{}^I$ and $C_{\mu\nu\rho}{}^I$ which become massive. After the absorption, $C_{\mu\nu\rho}{}^I$ becomes no longer auxiliary, but starts propagating as a massive scalar field. We fix all non-trivial cubic interactions in the total lagrangian, and quadratic interactions in all field equations. The superpartner fermion $\chi^I$ acquires a Dirac mass shared with the gaugino $\lambda^I$. As an independent confirmation, we give the superspace re-formulation of the component results.

PACS: 11.15.-q, 11.30.Pb, 12.60.Jv

[1)] E-Mail: H.Nishino@csulb.edu
[2)] E-Mail: Subhash.Rajpoot@csulb.edu

## 1. Introduction

The so-called Proca-Stueckelberg formalism was presented in 1930's [1][2] as a way to give masses to vector bosons. The mechanism utilizes a non-linear representation of an Abelian gauge transformation. The mass term is made gauge invariant by coupling a massless gauge boson to a real scalar field transforming non-linearly, which in the physical gauge is absorbed as the longitudinal component of the massive vector. The relevance of the Stueckelberg mechanism lies in the fact that it provides an alternative mechanism to the Higgs mechanism [3] to achieve gauge boson masses without spoiling renormalizability. It is to be noted that Proca-Stueckelberg mechanism [1] and the Higgs mechanism [3] are distinct, since the former mechanism only needs a single real scalar which is absorbed by the gauge boson to acquire a mass with no degrees of freedom left over.

It is the natural next step to *supersymmetrize* Stueckelberg formalism for *non-Abelian* gauge groups. For example, supersymmetric *Abelian* Stueckelberg formalism was presented [4] and applied to the minimal supersymmetric standard model (MSSM) [5]. Moreover, *non-Abelian* Stueckelberg formalism[3] was already formulated in superspace [6] in a somewhat disguised form. In the formalism of [6], both the vector multiplet and chiral multiplet within a single scalar superfield $V$ were used. The chiral multiplet was absorbed into the vector multiplet as the compensator multiplet, making the former massive. Afterwards, even $N=2$ supersymmetric non-Abelian Stueckelberg formalism was also formulated in [7], and this formulation was used for one-loop effective action [4]. Despite these developments in superspace, one does not yet have an explicit *component* formulation for $N=1$ supersymmetric non-Abelian YM theory.

Independent of these developments, we have presented in our previous paper [8] a supersymmetric non-Abelian Proca-Stueckelberg formalism in 3D. In the formulation in [8], the scalar compensator multiplet $(\varphi^I, \chi^I)$ separate from the vector multiplet $(A_\mu{}^I, \lambda^I)$ was used, where the scalar $\varphi^I$ is absorbed into the longitudinal component of $A_\mu{}^I$, making the latter massive. We were not aware whether the 4D analog of this formalism was possible at that time, except for those superspace results in [6][7][4].

In our more recent paper [9], we have presented a similar model for a supersymmetric non-Abelian tensor multiplet. Our field content in [9] was a non-Abelian Yang-Mills vector multiplet $(A_\mu{}^I, \lambda^I)$, a non-Abelian tensor multiplet $(B_{\mu\nu}{}^I, \chi^I, \varphi^I)$, and a compensator vector multiplet $(C_\mu{}^I, \rho^I)$. The $\varphi^I$ and $C_\mu{}^I$-fields are the compensator scalar and vector fields, respectively absorbed into $A_\mu{}^I$ and $B_{\mu\nu}{}^I$-fields. This formulation is further generalized to higher-order terms and more general representations of non-Abelian group in [10].

---

[3] Even though the original Stueckelberg formalism was only for the $U(1)$ Abelian group, we call this 'Stueckelberg formalism for *non-Abelian* groups'.

In the present paper, we present a formulation which is different from the superspace formulation [6], but in a direction similar to [8]. Our mechanism contains both the compensator scalar $\varphi^I$ and 2-form tensor $B_{\mu\nu}{}^I$, respectively absorbed into the YM $A_\mu{}^I$ and the 3-form non-Abelian tensor $C_{\mu\nu\rho}{}^I$ in the vector multiplet. To be more specific, we use two separate multiplets: the usual YM multiplet $(A_\mu{}^I, \lambda^I, C_{\mu\nu\rho}{}^I)$ and the tensor multiplet $(B_{\mu\nu}{}^I, \chi^I, \varphi^I)$ that do *not* belong to a single scalar superfield $V$. The tensor field $C_{\mu\nu\rho}{}^I$ is originally auxiliary dual to the conventional auxiliary field $D^I$. The $\varphi^I$ and $B_{\mu\nu}{}^I$ are compensator fields, and will be absorbed into the longitudinal components of $A_\mu{}^I$ and $C_{\mu\nu\rho}{}^I$, respectively. After the absorptions, these fields become massive. In particular, even though $C_{\mu\nu\rho}{}^I$ is originally 'auxiliary', it starts propagating as a massive spin 0 after the absorption. The *on-shell* degrees of freedom (DOF) count as $A_\mu{}^I(2), \lambda^I(2), C_{\mu\nu\rho}{}^I(0), B_{\mu\nu}{}^I(1), \chi^I(2), \varphi^I(1)$. After the absorptions, the last three compensator fields disappear, and the *on-shell* DOF count as $A_\mu{}^I(3), \lambda^I(4), C_{\mu\nu\rho}{}^I(1)$ (Cf. Table 1 below).

| DOF before Absorptions | $A_\mu{}^I$ | $\lambda^I$ | $C_{\mu\nu\rho}{}^I$ | $B_{\mu\nu}{}^I$ | $\chi^I$ | $\varphi^I$ |
|---|---|---|---|---|---|---|
| On-Shell | 2 | 2 | 0 | 1 | 2 | 1 |
| Off-Shell | 3 | 4 | 1 | 3 | 4 | 1 |

| DOF after Absorptions | $A_\mu{}^I$ | $\lambda^I$ | $C_{\mu\nu\rho}{}^I$ | $B_{\mu\nu}{}^I$ | $\chi^I$ | $\varphi^I$ |
|---|---|---|---|---|---|---|
| On-Shell | 3 | 4 | 1 | 0 | 0 | 0 |
| Off-Shell | 4 | 8 | 4 | 0 | 0 | 0 |

Table 1: DOF for Our Component Fields

We mention that our variant vector multiplet $(A_\mu{}^I, \lambda^I, C_{\mu\nu\rho}{}^I)$ is similar to the three-form multiplet as a variant formulation for a scalar multiplet introduced in [11]. However, one can easily see that our formulation is much more sophisticated, reflecting the progress in more than 30 years. For example, in eq. (2.10) in [11], the general Bianchi identities (BIds) for general $(p+1)$-form superfield strengths are given *without* Chern-Simons modifications that have non-trivial structures in our superspace $\mathcal{F}$, $G$ and $H$-BIds, as will be seen in section 4.

The organization of our present paper is as follows. In the next section, we give preliminaries and notational clarifications for non-Abelian Proca-Stueckelberg formulation only for bosonic fields, before supersymmetrization. In section 3, we fix our lagrangian, supersymmetry, and field equations. In section 4, we re-confirm the validity of our system in superspace [12][13]. Concluding remarks will be given in section 5.

## 2. Preliminaries for Proca-Stueckelberg Formalism

We consider an arbitrary continuous non-Abelian Lie group $G$ with generators satisfying,

$$[T^I, T^J] = f^{IJK} T^K \ , \tag{2.1}$$

where $f^{IJK}$ is the structure constant of $G$. Consider the Yang-Mills (YM) gauge field $A_\mu \equiv A_\mu{}^I T^I$ for the gauge group $G$, with the field strength[4]

$$F_{\mu\nu} \equiv \partial_\mu A_\nu - \partial_\nu A_\mu + m[A_\mu, A_\nu] \ , \tag{2.2}$$

where $m$ is the non-Abelian gauge coupling constant with the dimension of mass.[5]

As in the non-Abelian Proca-Stueckelberg formalism [2], we need the compensator field $\varphi^I$ in the adjoint representation, which will be absorbed into the longitudinal component of $A_\mu{}^I$. The finite gauge transformations for these fields will be [2]

$$(e^\varphi)' = e^{-\Lambda} e^\varphi \ , \quad (e^{-\varphi})' = e^{-\varphi} e^\Lambda \ , \tag{2.3a}$$

$$A_\mu' = m^{-1} e^{-\Lambda} \partial_\mu e^\Lambda + e^{-\Lambda} A_\mu e^\Lambda \ , \tag{2.3b}$$

$$F_{\mu\nu}' = e^{-\Lambda} F_{\mu\nu} e^\Lambda \ , \tag{2.3c}$$

with the $x$-dependent finite local YM gauge transformation parameters $\Lambda \equiv \Lambda^I(x) T^I$.

We can now define the covariant derivative acting on $e^\varphi$ by [2]

$$D_\mu e^\varphi \equiv \partial_\mu e^\varphi + m A_\mu e^\varphi \ , \tag{2.4}$$

---

[4] We sometimes omit the adjoint index $I$ in order to save space. We use $\mu, \nu, \cdots = 0, 1, 2, 3$ for bosonic space-time coordinates.

[5] In this paper, we comply with mass dimensions used in superspace in [13]. Accordingly, we assign the physical engineering dimension $0$ (or $1/2$) to a bosonic (or fermionic) fundamental field.

transforming *covariantly* under (2.3):

$$(D_\mu e^\varphi)' = e^{-\Lambda}(D_\mu e^\varphi) \ . \tag{2.5}$$

The *covariant* field strength of $\varphi$ is defined by

$$P_\mu \equiv (D_\mu e^\varphi) e^{-\varphi} \ , \tag{2.6}$$

transforming as

$$P_\mu{}' = e^{-\Lambda} P_\mu e^\Lambda \ . \tag{2.7}$$

Therefore the most appropriate choice for a gauge-covariant kinetic term for the $\varphi$-field is $-(1/2)(P_a{}^I)^2$. Accordingly, it is convenient to have the arbitrary infinitesimal variation

$$\delta P_\mu = [D_\mu - P_\mu, (\delta e^\varphi) e^{-\varphi}] + m \, \delta A_\mu \ . \tag{2.8}$$

Relevantly, the Bianchi identity (BId) for $P_\mu$ is

$$D_{[\mu} P_{\nu]} = +\tfrac{1}{2} m F_{\mu\nu} + \tfrac{1}{2}[P_\mu, P_\nu] \ . \tag{2.9}$$

We can now understand the Proca-Stueckelberg mechanism by the lagrangian

$$\mathcal{L}_1(x) = -\tfrac{1}{4}(F_{\mu\nu}{}^I)^2 - \tfrac{1}{2}(P_\mu{}^I)^2 \ . \tag{2.10}$$

Now redefine the gauge field by

$$\widetilde{A}_\mu \equiv e^{-\varphi} A_\mu e^\varphi + m^{-1} e^{-\varphi}(\partial_\mu e^\varphi) = m^{-1} e^{-\varphi} P_\mu e^\varphi \ , \tag{2.11}$$

so that the new field $\widetilde{A}_\mu$ and its field strength do *not* transform [1][2]

$$\widetilde{A}_\mu{}' = \widetilde{A}_\mu \ , \quad \widetilde{F}_{\mu\nu}{}' = \widetilde{F}_{\mu\nu} \ . \tag{2.12}$$

Because of the inverse relationships

$$P_\mu = m \, e^\varphi \widetilde{A}_\mu \, e^{-\varphi} \ , \quad F_{\mu\nu} = e^\varphi \widetilde{F}_{\mu\nu} \, e^{-\varphi} \ , \tag{2.13}$$

the exponential factors $e^{\pm\varphi}$ entirely disappear in the lagrangian in terms of *tilded* quantities:

$$\mathcal{L}_1(x) = -\tfrac{1}{4}(\widetilde{F}_{\mu\nu}{}^I)^2 - \tfrac{1}{2}m^2(\widetilde{A}_\mu{}^I)^2 \ , \tag{2.14}$$

while the explicit mass term for the gauge field $\widetilde{A}_\mu{}^I$ emerges.

An equivalent result can be also seen at the field equation level. The $A_\mu{}^I$-field equation from (2.10) is[6]

$$D_\nu F_\mu{}^{\nu I} - m P_\mu{}^I \doteq 0 \ . \tag{2.15}$$

In terms of the *tilded* fields, this is equivalent to

$$\widetilde{D}_\nu \widetilde{F}_\mu{}^{\nu I} - m^2 \widetilde{A}_\mu{}^I \doteq 0 \ , \tag{2.16}$$

where $\widetilde{D}_\mu$ coincides with $D_\mu$ with $A_\mu{}^I$ replaced by $\widetilde{A}_\mu{}^I$.

A similar formulation is possible for the 2-form non-Abelian compensator tenor field $B_{\mu\nu}{}^I$ absorbed into the longitudinal component of the 3-form tensor $C_{\mu\nu\rho}{}^I$. This mechanism is the non-Abelian generalization of the Abelian case in [14]

We start with the lagrangian

$$\mathcal{L}_2 \equiv -\tfrac{1}{48}(H_{\mu\nu\rho\sigma}{}^I)^2 - \tfrac{1}{12}(G_{\mu\nu\rho}{}^I)^2 \ , \tag{2.17}$$

where the field strengths $G$ and $H$ are defined by

$$G_{\mu\nu\rho}{}^I \equiv +3 D_{[\mu} B_{\nu\rho]}{}^I + m\, C_{\mu\nu\rho}{}^I \ , \tag{2.18a}$$

$$H_{\mu\nu\rho\sigma}{}^I \equiv +4 D_{[\mu} C_{\nu\rho\sigma]}{}^I + 6 f^{IJK} F_{[\mu\nu}{}^J B_{\rho\sigma]}{}^K \ . \tag{2.18b}$$

The $C_{\mu\nu\rho}{}^I$-field equation is

$$\frac{\delta \mathcal{L}_2}{\delta C_{\mu\nu\rho}{}^I} = -\tfrac{1}{6} D_\sigma H^{\mu\nu\rho\sigma I} - \tfrac{1}{6} m\, G^{\mu\nu\rho I} = -\tfrac{1}{6}\left( D_\sigma \widetilde{H}^{\mu\nu\rho\sigma I} + m^2 \widetilde{C}^{\mu\nu\rho I} \right) \doteq 0 \ , \tag{2.19}$$

where

$$\widetilde{C}_{\mu\nu\rho}{}^I \equiv C_{\mu\nu\rho}{}^I + 3 m^{-1} D_{[\mu} B_{\nu\rho]}{}^I \ , \quad G_{\mu\nu\rho}{}^I = m\, \widetilde{C}_{\mu\nu\rho}{}^I \ ,$$

$$\widetilde{H}_{\mu\nu\rho\sigma}{}^I \equiv 4 D_{[\mu} \widetilde{C}_{\nu\rho\sigma]}{}^I = H_{\mu\nu\rho\sigma}{}^I \ . \tag{2.20}$$

Note that the $F \wedge B$-term in (2.18b) cancels the term arising from the commutator on $B$, yielding exactly the same value both for $H_{\mu\nu\rho\sigma}{}^I$ and $\widetilde{H}_{\mu\nu\rho\sigma}{}^I$. The important point here is that this property is valid not only for Abelian case [14], but also for the present non-Abelian case. If we define $V_\mu{}^I$ by

$$V_\mu{}^I \equiv + \tfrac{1}{6} \epsilon_\mu{}^{\rho\sigma\tau} \widetilde{C}_{\rho\sigma\tau}{}^I \quad , \quad \widetilde{C}_{\mu\nu\rho}{}^I = +\epsilon_{\mu\nu\rho}{}^\sigma V_\sigma{}^I \quad , \tag{2.21}$$

then the original lagrangian $\mathcal{L}_2$ is re-expressed as

$$\mathcal{L}_2 = +(D_\mu V^{\mu I})^2 + \tfrac{1}{2} m^2 (V_\mu{}^I)^2 \quad , \tag{2.22}$$

---

[6] We use the symbol $\doteq$ for an equality that holds up to field equations.

If we vary this lagrangian by $V_\mu{}^I$, we get

$$D_\mu D_\nu V^{\nu I} - m^2 V^{\mu I} \doteq 0 \quad . \tag{2.23}$$

If $m \neq 0$, this field equation can be solved for $V_\mu{}^I$ as

$$V_\mu{}^I \doteq + m^{-1} D_\mu \phi^I \qquad (\phi^I \equiv +m^{-1} D_\mu V^{\mu I}) \quad . \tag{2.24}$$

We can re-express $\widetilde{H}_{\mu\nu\rho\sigma}{}^I$ and $\widetilde{C}_{\mu\nu\rho}{}^I$ in terms of $\phi^I$ as

$$\widetilde{H}_{\mu\nu\rho\sigma}{}^I \doteq -m^{-1} \epsilon_{\mu\nu\rho\sigma} D_\tau^2 \phi^I \quad , \quad \widetilde{C}_{\mu\nu\rho}{}^I \doteq + m^{-1} \epsilon_{\mu\nu\rho}{}^\sigma D_\sigma \phi^I \quad , \tag{2.25}$$

Using these in the original field equation (2.19), we get

$$D_\sigma \left( D_\tau^2 \phi^I - m^2 \phi^I \right) \doteq 0 \quad . \tag{2.26}$$

Here, the overall covariant derivative can be removed, under the ordinary boundary condition $\phi^I \to 0$ as $|x^i| \to \infty$, because the integration constant for the inside of the parentheses in (2.26) is to vanish, yielding the Klein-Gordon equation. This means nothing but the original system of $H$ and $G$ in $\mathcal{L}_2$ equivalent to a massive scalar field.

Even though this formulation seems just parallel to the Abelian case [14], the above formulation is valid also for *non-Abelian* case with non-trivial interactions.

The tensor fields $B_{\mu\nu}{}^I$ and $C_{\mu\nu\rho}{}^I$ have their own local 'tensorial' transformations with respect to their indices, such as $\delta_\beta B_{\mu\nu}{}^I = +2D_{[\mu} \beta_{\nu]}{}^I$ and $\delta_\gamma C_{\mu\nu\rho}{}^I = +3D_{[\mu} \gamma_{\nu\rho]}{}^I$. To be consistent with their field strengths, their complete forms are, for $\delta_\beta$

$$\delta_\beta B_{\mu\nu}{}^I = +2D_{[\mu} \beta_{\nu]}{}^I \quad , \tag{2.27a}$$

$$\delta_\beta C_{\mu\nu\rho}{}^I = +3f^{IJK} \beta_{[\mu}{}^J F_{\nu\rho]}{}^K \quad , \tag{2.27b}$$

and for $\delta_\gamma$

$$\delta_\gamma C_{\mu\nu\rho}{}^I = +3D_{[\mu}\gamma_{\nu\rho]}{}^I \quad , \tag{2.28a}$$

$$\delta_\gamma B_{\mu\nu}{}^I = -m\gamma_{\mu\nu}{}^I \quad , \tag{2.28b}$$

while $\delta_\beta A_\mu{}^I = \delta_\gamma A_\mu{}^I = 0$. It is not too difficult to confirm the invariances $\delta_\beta G_{\mu\nu\rho}{}^I = \delta_\beta H_{\mu\nu\rho\sigma}{}^I = \delta_\gamma G_{\mu\nu\rho}{}^I = \delta_\gamma H_{\mu\nu\rho\sigma}{}^I = 0$.

## 3. Lagrangian, Supersymmetry and Field Equations

Our field content is a VM $(A_\mu{}^I, \lambda^I, C_{\mu\nu\rho}{}^I)$ and a tensor multiplet $(B_{\mu\nu}{}^I, \chi^I, \varphi^I)$. Since we have understood the right kinetic terms for $\varphi^I$, $C_{\mu\nu\rho}{}^I$ and $B_{\mu\nu}{}^I$, it is easier to proceed for their supersymmetrization. Our action $I \equiv \int d^4x\, m^2 \mathcal{L}$ has the lagrangian[7]

$$\begin{aligned}\mathcal{L} =\ & -\tfrac{1}{4}(\mathcal{F}_{\mu\nu}{}^I)^2 + \tfrac{1}{2}(\overline\lambda^I \slashed{D}\lambda^I) - \tfrac{1}{48}(H_{\mu\nu\rho\sigma}{}^I)^2 \\ & -\tfrac{1}{12}(G_{\mu\nu\rho}{}^I)^2 + \tfrac{1}{2}(\overline\chi^I \slashed{D}\chi^I) - \tfrac{1}{2}(P_\mu{}^I)^2 + m(\overline\lambda^I\chi^I) \\ & + \tfrac{1}{24}f^{IJK}(\overline\lambda^I\gamma^{\mu\nu\rho\sigma}\chi^J)H_{\mu\nu\rho\sigma}{}^K - \tfrac{1}{2}f^{IJK}(\overline\lambda^J\gamma^\mu\lambda^J)P_\mu{}^K \quad .\end{aligned} \tag{3.1}$$

The field strengths $\mathcal{F}$, $G$, $H$ and $P$ are defined by

$$\begin{aligned}\mathcal{F}_{\mu\nu}{}^I & \equiv +2\partial_{[\mu}A_{\nu]}{}^I + mf^{IJK}A_\mu{}^J A_\nu{}^K + m^{-1}f^{IJK}P_\mu{}^J P_\nu{}^K \\ & \equiv +F_{\mu\nu}{}^I + m^{-1}f^{IJK}P_\mu{}^J P_\nu{}^K \quad ,\end{aligned} \tag{3.2a}$$

$$G_{\mu\nu\rho}{}^I \equiv +3D_{[\mu}B_{\nu\rho]}{}^I + mC_{\mu\nu\rho}{}^I \quad , \tag{3.2b}$$

$$H_{\mu\nu\rho\sigma}{}^I \equiv +4D_{[\mu}C_{\nu\rho\sigma]}{}^I + 6f^{IJK}F_{[\mu\nu}{}^J B_{\rho\sigma]}{}^K \quad , \tag{3.2c}$$

$$P_\mu{}^I \equiv \left[(\partial_\mu e^\varphi)\, e^{-\varphi} + mA_\mu\right]^I \equiv \left[(D_\mu e^\varphi)\, e^{-\varphi}\right]^I \quad . \tag{3.2d}$$

The field strengths $P$, $G$ and $H$ are the same as in section 2, while the new field strength $\mathcal{F}$ shifted from the original $F$ is to absorb $P^2$-terms arising frequently in our system.

Our action $I$ is invariant under global $N=1$ supersymmetry

$$\delta_Q A_\mu{}^I = +(\overline\epsilon\gamma_\mu\lambda^I) - m^{-1}f^{IJK}(\overline\epsilon\chi^J)P_\mu{}^K \quad , \tag{3.3a}$$

---

[7] We follow the dimensional assignments for fields in [13]. For example, all fundamental bosonic (or fermionic) fields have dimension $0$ (or $1/2$). This is the reason we need the overall factor $m^2$ in our action in front of the lagrangian.

$$\delta_Q \lambda^I = + \tfrac{1}{2}(\gamma^{\mu\nu}\epsilon)\mathcal{F}_{\mu\nu}{}^I - \tfrac{1}{24}(\gamma^{\mu\nu\rho\sigma}\epsilon)H_{\mu\nu\rho\sigma}{}^I$$
$$- \tfrac{1}{4}f^{IJK}\epsilon(\overline{\lambda}^J\chi^K) + \tfrac{1}{4}f^{IJK}(\gamma^\mu\epsilon)(\overline{\lambda}^J\gamma_\mu\chi^K) + \tfrac{1}{8}f^{IJK}(\gamma^{\mu\nu}\epsilon)(\overline{\lambda}^J\gamma_{\mu\nu}\chi^K)$$
$$+ \tfrac{1}{4}f^{IJK}(\gamma_5\gamma^\mu\epsilon)(\overline{\lambda}^J\gamma_5\gamma_\mu\chi^K) + \tfrac{3}{4}f^{IJK}(\gamma_5\epsilon)(\overline{\lambda}^J\gamma_5\chi^K) ~~, \tag{3.3b}$$

$$\delta_Q C_{\mu\nu\rho}{}^I = +(\overline{\epsilon}\gamma_{\mu\nu\rho}\chi^I) - 3f^{IJK}(\delta_Q A_{[\mu|}{}^J)B_{|\nu\rho]}{}^K ~~, \tag{3.3c}$$

$$\delta_Q B_{\mu\nu}{}^I = +(\overline{\epsilon}\gamma_{\mu\nu}\chi^I) ~~, \tag{3.3d}$$

$$\delta_Q \chi^I = + \tfrac{1}{6}(\gamma^{\mu\nu\rho}\epsilon)G_{\mu\nu\rho}{}^I - (\gamma^\mu\epsilon)P_\mu{}^I ~~, \tag{3.3e}$$

$$\left[(\delta_Q e^\varphi)e^{-\varphi}\right]^I = +(\overline{\epsilon}\chi^I) ~~. \tag{3.3f}$$

A useful lemma is the general variation of the field strengths

$$\delta\mathcal{F}_{\mu\nu}{}^I = +2D_{[\mu}(\delta A_{\nu]}{}^I) + 2m^{-1}f^{IJK}(\delta P_{[\mu}{}^J)P_{\nu]}{}^K ~~, \tag{3.4a}$$

$$\delta G_{\mu\nu\rho}{}^I = +3D_{[\mu}(\delta B_{\nu\rho]}{}^I) + m\left[\delta C_{\mu\nu\rho}{}^I + 3f^{IJK}(\delta A_{[\mu|}{}^J)B_{|\nu\rho]}{}^K\right] ~~, \tag{3.4b}$$

$$\delta H_{\mu\nu\rho\sigma}{}^I = +4D_{[\mu|}\left[\delta C_{|\mu\nu\rho]}{}^I + 3f^{IJK}(\delta A_{|\nu|}{}^J)B_{|\rho\sigma]}{}^K\right]$$
$$+ 4f^{IJK}(\delta A_{[\mu|}{}^J)G_{|\nu\rho\sigma]}{}^K - 6f^{IJK}(\delta B_{[\mu\nu|}{}^J)F_{|\rho\sigma]}{}^K ~~, \tag{3.4c}$$

$$\delta P_\mu{}^I = +D_\mu\left[(\delta e^\varphi)e^{-\varphi}\right]^I + f^{IJK}\left[(\delta e^\varphi)e^{-\varphi}\right]^J P_\mu{}^K + m\,\delta A_\mu{}^I ~~. \tag{3.4d}$$

Needless to say, these equations are general enough to be applied to supersymmetric variation $\delta_Q$ for each field. In particular, because of the second term in (3.3c), the $(\delta A) \wedge B$-term in (3.4b) and (3.4c) are cancelled, leaving only the $(\overline{\epsilon}\gamma_{\mu\nu\rho}\chi^I)$-term.

The field equations for $\lambda^I$, $\chi^I$, $A_\mu{}^I$, $B_{\mu\nu}{}^I$, $C_{\mu\nu\rho}{}^I$ and $\varphi^I$ from our lagrangian (3.1) are

$$\frac{\delta\mathcal{L}}{\delta\overline{\lambda}^I} = + \slashed{D}\lambda^I + m\chi^I + \tfrac{1}{24}f^{IJK}(\gamma^{\mu\nu\rho\sigma}\chi^J)H_{\mu\nu\rho\sigma}{}^K - f^{IJK}(\gamma^\mu\lambda^J)P_\mu{}^K \doteq 0 ~~, \tag{3.5a}$$

$$\frac{\delta\mathcal{L}}{\delta\overline{\chi}^I} = + \slashed{D}\chi^I + m\lambda^I - \tfrac{1}{24}f^{IJK}(\gamma^{\mu\nu\rho\sigma}\lambda^J)H_{\mu\nu\rho\sigma}{}^K \doteq 0 ~~, \tag{3.5b}$$

$$\frac{\delta\mathcal{L}}{\delta A_\mu{}^I} = -D_\nu\mathcal{F}^{\mu\nu I} - mP^{\mu I} - f^{IJK}\mathcal{F}^{\mu\nu J}P_\nu{}^K + \tfrac{1}{6}f^{IJK}H^{\mu\nu\rho\sigma J}G_{\nu\rho\sigma}{}^K$$
$$- mf^{IJK}(\overline{\lambda}^J\gamma^\mu\lambda^K) - \tfrac{1}{2}mf^{IJK}(\overline{\chi}^J\gamma^\mu\chi^K) \doteq 0 ~~, \tag{3.5c}$$

$$\frac{\delta\mathcal{L}}{\delta B_{\mu\nu}{}^I} = +D_\rho G^{\mu\nu\rho I} + \tfrac{1}{2}f^{IJK}F_{\rho\sigma}{}^J H^{\mu\nu\rho\sigma K} \doteq 0 ~~, \tag{3.5d}$$

$$\frac{\delta\mathcal{L}}{\delta C_{\mu\nu\rho}{}^I} = +D_\sigma H^{\mu\nu\rho\sigma I} + mG^{\mu\nu\rho I} - f^{IJK}D_\sigma(\overline{\lambda}^J\gamma^{\mu\nu\rho}\chi^K) \doteq 0 ~~, \tag{3.5e}$$

$$\frac{\delta\mathcal{L}}{[(\delta e^\varphi)e^{-\varphi}]^I} = +D_\mu P^{\mu I} - mf^{IJK}(\overline{\lambda}^J\chi^K) \doteq 0 ~~, \tag{3.5f}$$

up to $\mathcal{O}(\phi^3)$ corrections. Note that the coefficients of the $(\overline{\lambda}^J\gamma^\mu\lambda^K)$-term in (3.5c) is twice as large as that of the $(\overline{\chi}^J\gamma^\mu\chi^K)$-term, due to the interaction $f^{IJK}(\overline{\lambda}^I\gamma^\mu\lambda^J)P_\mu{}^K$ in the lagrangian.

As we have described the mechanism in section 2, the physical significance of these field equations is clear. First, the original scalar field $\varphi^I$ is absorbed into the longitudinal component of $A_\mu{}^I$ making the latter massive. Second, the tensor field $B_{\mu\nu}{}^I$ plays the role of compensator absorbed into the longitudinal component of $C_{\mu\nu\rho}{}^I$, making the latter massive. Third, this sort of compensator mechanisms should be also consistent with supersymmetry. In fact, the original $\chi^I$ and $\lambda^I$-fields form a massive Dirac field with the common mass $m$. As given in Table 1, the counting of DOF also works, consistently with supersymmetry.

We can confirm also the mutual consistency among our field equations (3.5). For example, we can apply the divergence operation on (3.5c), (3.5d) and (3.5e) to see, if they vanish upon the use of other field equations. For example, the $D_\rho$-operation on (3.5e) yields

$$\begin{aligned}
0 \overset{?}{=} &+ D_\rho \left[ +D_\sigma H^{\mu\nu\rho\sigma\,I} + mG^{\mu\nu\rho\,I} - f^{IJK}D_\sigma(\overline{\lambda}^J\gamma^{\mu\nu\rho\sigma}\chi^K) \right] \\
= &+ \tfrac{1}{2} mf^{IJK} F_{\rho\sigma}{}^J H^{\mu\nu\rho\sigma\,K} + mD_\rho G^{\mu\nu\rho\,I} + \mathcal{O}(\phi^3) \\
\doteq &+ \tfrac{1}{2} mf^{IJK} F_{\rho\sigma}{}^J H^{\mu\nu\rho\sigma\,K} + m\left[ -\tfrac{1}{2} f^{IJK} F_{\sigma\tau}{}^J H^{\mu\nu\sigma\tau\,K} \right] + \mathcal{O}(\phi^3) \\
= &+\mathcal{O}(\phi^3) \quad (Q.E.D.)
\end{aligned} \qquad (3.6)$$

This tells us, e.g., why the term $mG$ is needed in the $C$-field equation. If it were not there, the $mfFH$-term would *not* be cancelled. Similar confirmation can be done also for $A$ and $B$-field equations. This confirmation provides the important consistency check for non-Abelian tensor, in particular, the non-trivial couplings of $B_{\mu\nu}{}^I$ and $C_{\mu\nu\rho}{}^I$-fields.

## 4. Superspace Re-Confirmation

Based on the prescription for the purely bosonic system in section 2, we can develop the superspace formulation for $N = 1$ supersymmetric non-Abelian Proca-Stueckelberg formalism. The superfields we need are $(A_a{}^I, \lambda_\alpha{}^I, C_{abcd}{}^I)$[8] for the vector multiplet, and $(B_{ab}{}^I, \chi_\alpha{}^I, \varphi^I)$ for the compensator tensor multiplet.

---

[8] We follow the notation in [13][12]. Namely, we use the indices $A = (a,\alpha)$, $B = (b,\beta)$, $\cdots$ for the local Lorentz coordinates in superspace, where $a, b, \cdots = (0), (1), (2), (3)$ are for the bosonic coordinates, while $\alpha, \beta, \cdots = 1, 2, 3, 4$ are for fermionic coordinates. Note that our (anti)symmetrization convention is such as $M_{[AB)} \equiv M_{AB} - (-)^{AB} M_{BA}$ *without* the factor of $1/2$ [13].

In superspace, $P_a{}^I$ in (2.6) is generalized to

$$P_A{}^I \equiv \left[ (\nabla_A e^\varphi) e^{-\varphi} \right]^I . \qquad (4.1)$$

The $\varphi \equiv \varphi(Z)$ is now a scalar superfield whose $\theta = 0$ component is the component field $\varphi(x)$.

In our superspace, we need an additional superfield $L_{ABC}$ whose non-vanishing component is $L_{\alpha\beta c} = +2(\gamma_c)_{\alpha\beta}$. This superfield is very similar to the corresponding one in our 4D formulation [14]. There are seven superfield strengths $\mathcal{F}_{AB}{}^I$, $G_{ABC}{}^I$, $H_{ABCD}{}^I$, $L_{ABC}$, $P_A{}^I$, $T_{AB}{}^C$ and $R_{AB}{}^{cd}$, where the first four superfields are respectively defined in terms of potential superfields, $A_A{}^I$, $B_{AB}{}^I$, $C_{ABC}{}^I$ and $M_{AB}$ by 'rotation' operations, together with non-trivial Chern-Simons terms:

$$\mathcal{F}_{AB}{}^I \equiv + \nabla_{[A} A_{B)}{}^I + m f^{IJK} A_A{}^J A_B{}^K - T_{AB}{}^C A_C{}^I + m^{-1} f^{IJK} P_A{}^J P_B{}^K$$
$$\equiv + F_{AB}{}^I + m^{-1} f^{IJK} P_A{}^J P_B{}^K \qquad (4.2a)$$

$$G_{ABC}{}^I \equiv + \tfrac{1}{2!} \nabla_{[A} B_{BC)}{}^I - \tfrac{1}{2!} T_{[AB}{}^D B_{D|C)}{}^I + m C_{ABC}{}^I , \qquad (4.2b)$$

$$H_{ABCD}{}^I \equiv + \tfrac{1}{3!} \nabla_{[A} C_{BCD)}{}^I - \tfrac{1}{(2!)^2} T_{[AB|}{}^E C_{E|CD)}{}^I$$
$$+ \tfrac{1}{(2!)^2} f^{IJK} F_{[AB}{}^J B_{CD)}{}^K - \tfrac{1}{3!} m^{-1} L_{[ABC} P_{D)}{}^I , \qquad (4.2c)$$

$$L_{ABC} \equiv + \tfrac{1}{2!} \nabla_{[A} M_{BC)} - \tfrac{1}{2!} T_{[AB|}{}^D M_{D|C)} , \qquad (4.2d)$$

while $P_A{}^I$ has been already defined by (4.1).

The BIds for these field strengths can be relatively easily obtained by going from the local Lorentz frame to *curved* coordinate frame, in order to eliminate super-torsion dependent terms:

$$\mathcal{F}_{MN}{}^I \equiv + \partial_{[M} A_{N)}{}^I + m f^{IJK} A_M{}^J A_N{}^K + m^{-1} f^{IJK} P_M{}^J P_N{}^K$$
$$\equiv + F_{MN}{}^I + m^{-1} f^{IJK} P_M{}^J P_N{}^K \qquad (4.3a)$$

$$G_{MNP}{}^I \equiv + \tfrac{1}{2!} \nabla_{[M} B_{NP)}{}^I + m C_{MNP}{}^I , \qquad (4.3b)$$

$$H_{MNPQ}{}^I \equiv + \tfrac{1}{3!} \nabla_{[M} C_{NPQ)}{}^I + \tfrac{1}{(2!)^2} f^{IJK} F_{[MN}{}^J B_{PQ)}{}^K - \tfrac{1}{3!} m^{-1} L_{[MNP} P_{Q)}{}^I , \qquad (4.3c)$$

$$P_M{}^I \equiv \left[ (\nabla_M e^\varphi) e^{-\varphi} \right]^I , \qquad (4.3d)$$

$$L_{MNP} \equiv + \tfrac{1}{2!} \nabla_{[M} M_{NP)} . \qquad (4.3e)$$

Eqs. (4.3a) through (4.3d) are the superspace generalizations of our component results in (3.2). By applying the superspace rotation operations on (4.3), we get the BIds

$$+ \tfrac{1}{2!} \nabla_{[M} \mathcal{F}_{NP)}{}^I - \tfrac{1}{2!} f^{IJK} \mathcal{F}_{[MN|}{}^J P_{|P)}{}^K \equiv 0 ~, \tag{4.4a}$$

$$+ \tfrac{1}{3!} \nabla_{[M} G_{NPQ)}{}^I - m H_{MNPQ}{}^I - \tfrac{1}{3!} L_{[MNP} P_{Q)}{}^I \equiv 0 ~, \tag{4.4b}$$

$$+ \tfrac{1}{4!} \nabla_{[M} H_{NPQR)}{}^I - \tfrac{1}{(2!)(3!)} f^{IJK} F_{[MN|}{}^J G_{|PQR)}{}^K - \tfrac{1}{(3!)(2!)} L_{[MNP} \mathcal{F}_{QR)}{}^I \equiv 0 ~, \tag{4.4c}$$

$$+ \nabla_{[M} P_{N)}{}^I - m \mathcal{F}_{MN}{}^I \equiv 0 ~, \tag{4.4d}$$

$$+ \tfrac{1}{3!} \nabla_{[M} L_{NPQ)} \equiv 0 ~. \tag{4.4e}$$

These curved-index BIds are rewritten in terms of local-Lorentz-index BIds, where the supertorsion-dependent terms are recovered. After all, the superfield strengths $\mathcal{F}_{AB}{}^I$, $G_{ABC}{}^I$, $H_{ABC}{}^I$, $P_A{}^I$, $L_{ABC}$, $T_{AB}{}^C$ and $R_{AB}{}^{cd}$ satisfy the BIds:

$$+ \tfrac{1}{2!} \nabla_{[A} \mathcal{F}_{BC)}{}^I - \tfrac{1}{2!} T_{[AB|}{}^D \mathcal{F}_{D|C)}{}^I - \tfrac{1}{2!} f^{IJK} \mathcal{F}_{[AB|}{}^J P_{|C)}{}^K \equiv 0 ~, \tag{4.5a}$$

$$+ \tfrac{1}{3!} \nabla_{[A} G_{BCD)}{}^I - \tfrac{1}{(2!)^2} T_{[AB|}{}^E G_{E|CD)}{}^I - m H_{ABCD}{}^I - \tfrac{1}{3!} L_{[ABC} P_{D)}{}^I \equiv 0 ~, \tag{4.5b}$$

$$+ \tfrac{1}{4!} \nabla_{[A} H_{BCDE)}{}^I - \tfrac{1}{(2!)(3!)} T_{[AB|}{}^F H_{F|CDE)}{}^I$$
$$- \tfrac{1}{(2!)(3!)} f^{IJK} F_{[AB|}{}^J G_{|CDE)}{}^K - \tfrac{1}{(3!)(2!)} L_{[ABC} \mathcal{F}_{DE)}{}^I \equiv 0 ~, \tag{4.5c}$$

$$+ \nabla_{[A} P_{B)}{}^I - T_{AB}{}^C P_C{}^I - m \mathcal{F}_{AB}{}^I \equiv 0 ~, \tag{4.5d}$$

$$+ \tfrac{1}{3!} \nabla_{[A} L_{BCD)} - \tfrac{1}{(2!)^2} T_{[AB|}{}^E L_{E|CD)} \equiv 0 ~, \tag{4.5e}$$

$$+ \tfrac{1}{2!} \nabla_{[A} T_{BC)}{}^D - \tfrac{1}{2!} T_{[AB|}{}^E T_{E|C)}{}^D - \tfrac{1}{2(2!)} R_{[AB|\,e}{}^f (\mathcal{M}_f{}^e)_{|C)}{}^D \equiv 0 ~, \tag{4.5f}$$

where $(\mathcal{M}_a{}^b)_C{}^D$ is the Lorentz generator.

For readers who are not yet convinced of the total consistency, we give the additional confirmation of each of our new BIds (4.4a) through (4.4e) by taking their superspace rotations:

**(i)** The $\mathcal{F}$-BId: In this case, the superspace rotation of (4.4a) will be

$$0 \stackrel{?}{=} \nabla_{[M|} \left[ \nabla_{|N|} \mathcal{F}_{|PQ|}{}^I - f^{IJK} \mathcal{F}_{|NP|}{}^J P_{|Q)}{}^K \right]$$
$$= + \tfrac{1}{2} m f^{IJK} F_{[MN|}{}^J \mathcal{F}_{|PQ)}{}^K - f^{IJK} (\nabla_{[M} \mathcal{F}_{|NP|}{}^J) P_{|Q)}{}^K - f^{IJK} \mathcal{F}_{[MN|}{}^J (\nabla_{|P|} P_{|Q)}{}^K) ~. \tag{4.6}$$

For the first term, we rewrite $F$ by $\mathcal{F} - (1/2)[P, P\}$, for the second term we use the original $\mathcal{F}$-BId (4.4a), while for the last term we use the $P$-BId (4.4d). We next see that the term $\mathcal{F} \wedge \mathcal{F}$-term disappears due to the (anti)symmetries of indices $_{[MNPQ)}$ under $f^{IJK}$, while the $\mathcal{F} \wedge P \wedge P$-terms are combined to form the coefficient $f^{I[K|J} f^{J|LM]} \equiv 0$ vanishing due to the Jacobi identity.

**(ii)** The $B$-BId: Similarly, we get the superspace rotation of (4.4b) as

$$0 \stackrel{?}{=} \nabla_{[M|} \left[ 4\nabla_{|N|} G_{|PQR)}{}^I - m H_{|NPQR)}{}^I - 4 L_{|NPQ|} P_{|R)}{}^I \right]$$
$$= +2m f^{IJK} F_{[MN|}{}^J G_{|PQR)}{}^K - m \nabla_{[M} H_{NPQR)}{}^I$$
$$- 4(\nabla_{[M|} L_{|NPQ|}) P_{|R)}{}^I + 4 L_{[MNP|} \nabla_{|Q|} P_{|R)}{}^I \quad (4.7)$$

We use the $H$, $L$ and $P$-BIds for the second, third and forth terms, respectively. There then arise $F \wedge G$ and $L \wedge F$-terms, both of which cancel themselves. The fact that $F \wedge G$-term in the $H$-BId (4.4c) is with $F$ instead of $\mathcal{F}$ also plays a crucial role in the cancellation here.

**(iii)** $H$-BId: The superspace rotation of (4.4c) is

$$0 \stackrel{?}{=} \nabla_{[M} \left[ 5 \nabla_{|N|} H_{|PQRS)}{}^I - 10 f^{IJK} F_{|NP|}{}^J G_{|QRS)}{}^K - 10 L_{|NPQ|} \mathcal{F}_{RS)}{}^I \right]$$
$$= + \tfrac{5}{2} m f^{IJK} F_{[MN|}{}^J H_{|PQRS)}{}^K - 10 f^{IJK} F_{[MN|}{}^J \nabla_{|P|} G_{|QRS)}{}^K$$
$$- 10 \left( \nabla_{[M|} L_{|NPQ|} \right) \mathcal{F}_{|RS)}{}^I + 10 L_{[MNP|} \nabla_{|Q|} \mathcal{F}_{|RS)}{}^I . \quad (4.8)$$

We use the $G$, $L$-BIds and $\mathcal{F}$-BIds for the second, third, and last terms, respectively. After this, there arise the $F \wedge H$ and $L \wedge F \wedge P$ or $L \wedge \mathcal{F} \wedge P$-terms. The former cancel themselves, while the latter two terms also cancel each other as

$$0 \stackrel{?}{=} +10 f^{IJK} L_{[MNP|}{}^J \left( \mathcal{F}_{|QR|}{}^J - F_{|QR|}{}^J \right) P_{|S)}{}^K$$
$$= - \tfrac{5}{3} m f^{I[K|J} f^{J|LM]} L_{[MNP|} P_{|Q|}{}^L P_{|R|}{}^M P_{|S)}{}^K \equiv 0 , \quad (4.9)$$

due to the Jacobi identity $f^{I[K|N} f^{N|LM]} \equiv 0$. In other words, the difference between $\mathcal{F}$ and $F$ in the $L \wedge F \wedge P$ and $L \wedge \mathcal{F} \wedge P$-terms does *not* matter after all in (4.9).

**(iv)** The $P$-BId: This is simply obtained as

$$\nabla_{[M} P_{N)}{}^I = \nabla_{[M} \left[ (\nabla_{N)} e^{\varphi}) e^{-\varphi} \right]$$
$$= \left[ (\nabla_{[M} \nabla_{N)} e^{\varphi}) e^{-\varphi} \right]^I - \left[ (\nabla_{[M|} e^{\varphi})(\nabla_{|N)} e^{-\varphi}) \right]^I$$
$$= m F_{MN}{}^I + [P_M, P_N\}^I , \quad (4.10)$$

where the only subtlety is the operation $\nabla_{[M} \nabla_{N)} e^\varphi = m F_{MN} e^\varphi$. This is due to the special property of the compensator superfield $\varphi$. Eq. (4.10) is also the superspace generalization of the bosonic-component BId (2.9).

**(v)** The $L$-BId: This case is almost trivial, so that we do not go into the details.

All of these consistencies imply the following important and non-trivial facts:

(1) All the non-trivial Chern-Simons-type terms in our BIds in (4.5) have been confirmed in an extremely non-trivial manner.

(2) In particular, the non-trivial fact that the $F \wedge G$-term with $F$, while the $L \wedge \mathcal{F}$-term with $\mathcal{F}$ in the $H$-BId (4.5c) has been confirmed.

We mention the well-known presentation by M. Muller [15] about the 2-form construction for a tensor multiplet. Similar to the aforementioned ref. [11], our formulation has differences as well as similarities, compared with [15]. The most important difference is the presence of supersymmetric Chern-Simons terms in the $G$-superfield strengths (4.3b) or $G$-BIds (4.5b), reflecting non-trivial interaction structures of our Proca-Stueckelberg mechanism in superspace.

Some readers may wonder, why we do *not* use the prepotential formulation for the tensor multiplet [16][17][13], which would be simpler and straightforward. To that question, we repeat the same answer presented in section 4 of our previous paper [18]. Namely, the short answer is that there is *no* known consistent prepotential formulation for *non-Abelian* tensor multiplet. A long answer is summarized as follows:

(1) The main obstruction for prepotential for *non-Abelian* tensor multiplet pops up in the basic commutator (*not* anti-commutator) on the scalar prepotential $L$:

$$\left[ \nabla_\alpha, \overline{\nabla}_{\dot\beta} \right] L = c_1 \left( \sigma^{cde} \right)_{\alpha\dot\beta} G_{cde} + c_2 \, \mathrm{tr} \left( W_\alpha \overline{W}_{\dot\beta} \right) \ . \quad (4.11)$$

The problem is that the $G$-term on the right side is supposed to carry the adjoint index, while the second $W\overline{W}$-term does *not*, due to the trace on the adjoint index.

(2) One might expect that the already-established prepotential formulation [16][17][13] should be valid to any interactions, including *non-Abelian* ones. However, such a conjecture will *not* be realized, because our tensor multiplet carries an *adjoint index*, which is beyond the scope of the conventional prepotential [17].

These are the reasons why even the off-shell prepotential formulation for the Abelian tensor multiplet [16][17][13] *not* work in the *non-Abelian* case.

As usual, the next step is to satisfy the BIds in (4.5) by consistent constraints. We found the appropriate constraints at the engineering dimensions $d \leq 1$ are

$$T_{\alpha\beta}{}^c = +2(\gamma^c)_{\alpha\beta} \ , \quad L_{\alpha\beta c} = +2(\gamma_c)_{\alpha\beta} \ , \tag{4.12a}$$

$$F_{\alpha b}{}^I = -(\gamma_b \lambda^I)_\alpha \ , \quad G_{\alpha bc}{}^I = -(\gamma_{bc}\chi^I)_\alpha \ , \quad H_{\alpha bcd}{}^I = -(\gamma_{bcd}\lambda^I)_\alpha \ , \quad P_\alpha{}^I = -\chi_\alpha{}^I \ , \tag{4.12b}$$

$$\nabla_\alpha \chi_\beta{}^I = -(\gamma^c)_{\alpha\beta} P_c{}^I - \tfrac{1}{6}(\gamma^{cde})_{\alpha\beta} G_{bcd}{}^I \ , \tag{4.12c}$$

$$\begin{aligned}
\nabla_\alpha \lambda_\beta{}^I =\ & +\tfrac{1}{2}(\gamma^{cd})_{\alpha\gamma} F_{cd}{}^I + \tfrac{1}{24}(\gamma^{cdef})_{\alpha\beta} H_{cdef}{}^I \\
& + \tfrac{1}{4} C_{\alpha\beta} f^{IJK}(\overline{\lambda}^J \chi^K) + \tfrac{1}{4}(\gamma^c)_{\alpha\beta} f^{IJK}(\overline{\lambda}^J \gamma_c \chi^K) - \tfrac{1}{8}(\gamma^{cd})_{\alpha\beta} f^{IJK}(\overline{\lambda}^J \gamma_{cd} \chi^K) \\
& - \tfrac{1}{4}(\gamma_5 \gamma^c)_{\alpha\beta} f^{IJK}(\overline{\lambda}^J \gamma_5 \gamma_c \chi^K) - \tfrac{3}{4}(\gamma_5)_{\alpha\beta} f^{IJK}(\overline{\lambda}^J \gamma_5 \chi^K) \ .
\end{aligned} \tag{4.12d}$$

Corresponding to the component computation, these constraints are valid up to $\mathcal{O}(\phi^3)$-terms. All other independent components at $d \leq 1$, such as $G_{\alpha\beta\gamma}{}^I$, $R_{\alpha\beta c}{}^d$ or $L_{abc}$ are zero. It is not too difficult to confirm the satisfaction of all the BIds (4.5) at $d \leq 1$ by these constraints.[9]

As usual in superspace, the BIds at $d = 3/2$ yield the following relationships

$$\nabla_\alpha \mathcal{F}_{bc}{}^I = +(\gamma_{[b} \nabla_{c]} \lambda^I)_\alpha - f^{IJK}(\gamma_{[b|}\chi^J)_\alpha P_{|c]}{}^K + f^{IJK}\chi_\alpha{}^J \mathcal{F}_{bc}{}^K \ , \tag{4.13a}$$

$$\nabla_\alpha G_{bcd}{}^I = -\tfrac{1}{2}(\gamma_{[bc} \nabla_{|d]} \chi^I)_\alpha - m(\gamma_{bcd}\lambda^I)_\alpha \ , \tag{4.13b}$$

$$\nabla_\alpha H_{bcde}{}^I = +\tfrac{1}{6}(\gamma_{[bcd|}\nabla_{|e]}\chi^I)_\alpha - \tfrac{1}{6} f^{IJK}(\gamma_{[b|}\lambda^J)_\alpha G_{|cde]}{}^K + \tfrac{1}{4} f^{IJK}(\gamma_{[bc|}\chi^J)_\alpha \mathcal{F}_{|de]}{}^K \ , \tag{4.13c}$$

$$\nabla_\alpha P_b{}^I = +\nabla_b \chi_\alpha{}^I + m(\gamma_b \lambda^I)_\alpha + f^{IJK}\chi_\alpha{}^J P_b{}^K \ . \tag{4.13d}$$

These are consistent with the component results in (3.4).

These relationships will be used to get the bosonic superfield equations from fermionic ones. Note that the superspace constraints so far will *not* fix the fermionic superfield equations. The reason is that our vector and tensor multiplets are *off-shell* multiplets. In other

---

[9] The confirmation is valid up to cubic-order terms $\mathcal{O}(\phi^3)$. These cubic terms correspond to the quartic terms at the lagrangian level, which are usually omitted in supergravity system [19]. Even though our system is not supergravity system, such an analogy is legitimate, due to the presence of the constant $m^{-1}$ with the dimension of length like the gravitational constant $\kappa$. See [19] or [9] for similar treatments.

words, it is the component-lagrangian (3.1) that provides the fermionic field equations:

$$\Lambda_\alpha{}^I \equiv (\slashed{\nabla}\lambda^I)_\alpha + m\chi_\alpha{}^I - \tfrac{1}{24} f^{IJK}(\gamma^{bcde}\chi^J)_\alpha H_{bcde}{}^K - f^{IJK}(\gamma^b\lambda^J)_\alpha P_b{}^K \doteq 0 \ , \quad (4.14\text{a})$$

$$X_\alpha{}^I \equiv (\slashed{\nabla}\chi^I)_\alpha + m\lambda_\alpha{}^I - \tfrac{1}{24} f^{IJK}(\gamma^{bcde}\lambda^J)_\alpha H_{bcde}{}^K \doteq 0 \ . \quad (4.14\text{b})$$

As usual in superspace formulation, the application of fermionic derivatives multiplied by $\gamma$-matrices on these equations $\Lambda_\alpha{}^I \doteq 0$ and $X_\alpha{}^I \doteq 0$ yield the remaining bosonic $A$, $B$, $C$ and $\varphi$-superfield equations:

$$+ \tfrac{1}{4}(\gamma^a)^{\alpha\beta}\nabla_\beta\Lambda_\alpha{}^I \doteq + \nabla_b\mathcal{F}^{abI} + mP^{aI} - f^{IJK}\mathcal{F}^{abJ}P_b{}^K - \tfrac{1}{6} f^{IJK}H^{abcdJ}G_{bcd}{}^K$$

$$+ mf^{IJK}(\overline{\lambda}^J\gamma^a\lambda^K) + \tfrac{1}{2} mf^{IJK}(\overline{\chi}^J\gamma^a\chi^K) \doteq 0 \ , \quad (4.15\text{a})$$

$$+ \tfrac{1}{4}(\gamma^{ab})^{\alpha\beta}\nabla_\beta X_\alpha{}^I \doteq \nabla_c G^{abcI} + \tfrac{1}{2} f^{IJK} F_{cd}{}^J H^{abcdK} \doteq 0 \ , \quad (4.15\text{b})$$

$$- \tfrac{1}{4}(\gamma^{abc})^{\alpha\beta}\nabla_\beta\Lambda_\alpha{}^I \doteq + \nabla_d H^{abcdI} + mG^{abcI} - f^{IJK}\nabla_d(\overline{\lambda}^J\gamma^{abcd}\chi^K) \doteq 0 \ , \quad (4.15\text{c})$$

$$- \tfrac{1}{4}\nabla^\alpha X_\alpha{}^I \doteq + \nabla_a P^{aI} - mf^{IJK}(\overline{\lambda}^J\chi^K) \doteq 0 \ . \quad (4.15\text{d})$$

These are consistent with our bosonic component field equations (3.5) up to $\mathcal{O}(\phi^3)$-terms.

## 5. Concluding Remarks

In this paper, we have established the very economical but still non-trivial and consistent interactions for $N=1$ supersymmetric non-Abelian Proca-Stueckelberg mechanism in 4D, both in component and superspace languages. All we need are only two multiplets $(A_\mu{}^I, \lambda^I, C_{\mu\nu\rho}{}^I)$ and $(B_{\mu\nu}{}^I, \chi^I, \varphi^I)$. We have confirmed the basic consistency of the system, despite the non-trivial non-Abelian interactions. Even though the system has the coupling $m^{-1}$ with the inverse mass dimension for non-renormalizability, still non-trivial consistency with supersymmetry has been confirmed.

Compared with the recent progress in the similar direction, such as [10] in 4D or [8] in 3D, our system here is much simpler and economical. We do *not* need any extra vector multiplet to be absorbed into the original vector multiplet. Only two multiplets $(A_\mu{}^I, \lambda^I, C_{\mu\nu\rho}{}^I)$ and $(B_{\mu\nu}{}^I, \chi^I, \varphi^I)$ are enough. Still our system shows how a non-Abelian tensor works in terms of compensator mechanism. Our system has also a supersymmetric non-Abelian tensor system as a bonus. Our mechanism is supposed to be the simplest system of this kind at least in 4D.

As has been mentioned, the mass dimensions in our system are the same as in superspace [13]. For this reason, we need the special overall factor $m^2$ in the action $I \equiv \int d^4x\, m^2 \mathcal{L}$. This property is very similar to supergravity theories with the factor $\kappa^{-2}$ before their lagrangians [19]. Even though our system is globally supersymmetric, the similarity of our system to type IIB supergravity [19] arises from the involvement of the non-renormalizable coupling $m^{-1}$ with the dimension of inverse-mass similarly to the gravitational coupling $\kappa$. This property results in the non-renormalizable interactions, such as the Pauli-terms $\overline{\lambda}\chi H$ or $\overline{\lambda}\lambda P$ in our lagrangian. This aspect gives the justification for our lagrangian fixed up to quartic terms, while our field equations only up to cubic terms $\mathcal{O}(\phi^3)$. For example, in type IIB supergravity [19], all *quadratic fermionic terms* in field equations were omitted, due to impractical complications. In contrast, we have included all *quadratic fermionic terms* in field equations, so that our treatment is better than [19].

As the confirmation of the total system with supersymmetry, we have performed non-trivial cross-checks of our field equations (3.5). The total consistency among field equations has been re-confirmed by the divergence of the $A$, $B$ and $C$-field equations, *e.g.*, (3.6).

Very few results have been ever presented for *supersymmetric* Proca-Stueckelberg formalism for *non-Abelian* gauge group in 4D, except for [6][7][4]. There seem to be three main reasons. First, we need the non-renormalizable coupling $m^{-1}$ that seems unusual as globally supersymmetric theory. Second, if there is the coupling $m^{-1}$, this implies that the limit $m \to 0$ is *not* smooth. Therefore, we can *not* extrapolate the *un-gauged* case with $m = 0$ to the gauged case $m \neq 0$ so smoothly. This seems to be the reason why starting with an *un-gauged* sigma-model for a group manifold, and then going to the *gauged* sigma-model did not work in the past. Third, certain problems at quantum level have been known for Proca-Stueckelberg formalism at quantum level [20], which provides a disadvantage against such supersymmetrization, as will be mentioned below.

As non-trivial re-confirmation, we have established superspace reformulation. We have started with the definition of superfield strength (4.2), and next we have derived all BIds for our new superfield strengths $\mathcal{F}_{AB}{}^I$, $G_{ABC}{}^I$, $H_{ABCD}{}^I$, $P_A{}^I$ and $L_{ABC}$ in (4.4). The mutual consistency of these BIds have been also re-confirmed by taking their superspace rotations in (4.6) through (4.10). All component field equations have been also re-obtained in superspace.

Compared with the corresponding formalism in 3D [8], there are similarities as well as differences. Similarity is such that we have the Dirac mass term as the mixture of the gaugino field $\lambda$ and the super-partner $\chi$ of the compensator field $\varphi$. The most important difference is the existence of the $B_{\mu\nu}{}^I$ that complicates the system in 4D. But the advantage is that this field shows how the Proca-Stueckelberg mechanism works for supersymmetric non-Abelian tensors.

The success of our formulation in 4D as well as in 3D [8] is encouraging, and it is natural to expect that similar Proca-Stueckelberg formalisms are possible also in higher-dimensions $D \geq 5$, as long as there exists a multiplet involving a compensator scalar field.

As has been mentioned, the *non-supersymmetric* non-Abelian Proca-Stueckelberg formalism has certain problems at quantum level [20]. In this paper, we do not address ourselves to such problems. Nevertheless, we mention the general feature of supersymmetry, *i.e.,* the quantum behavior of supersymmetric system is much better, compared with the corresponding non-supersymmetric systems. In other words, it may well be the case that supersymmetric Proca-Stueckelberg formalism has an intrinsic solution to the conventional problem of Proca-Stueckelberg formalism at quantum level.

We are grateful to W. Siegel for intensive and important discussions. This work is supported in part by Department of Energy grant # DE-FG02-10ER41693.